\documentclass{article} 
\usepackage{iclr2027_conference,times}
\iclrfinalcopy

\usepackage{amsmath,amsfonts,bm}

\def\eqref#1{equation~\ref{#1}}

\def\1{\bm{1}}

\DeclareMathAlphabet{\mathsfit}{\encodingdefault}{\sfdefault}{m}{sl}
\SetMathAlphabet{\mathsfit}{bold}{\encodingdefault}{\sfdefault}{bx}{n}

\usepackage{enumitem} 
\usepackage{hyperref,url,booktabs,xcolor,graphicx,amsmath,amssymb,multirow,array}
\usepackage[T1]{fontenc}
\usepackage{wrapfig}
\usepackage{tabularx}
\hypersetup{hidelinks}

\title{TASTE: Taxonomy-Aware Symbolic Piano Generation from Textual Description}

\author{Mingyang Yao$^{1,2}$, Jingyue Huang$^{1}$, Ke Chen$^{1}$, Taylor Berg-Kirkpatrick$^{1}$, Shlomo Dubnov$^{1}$ \\
$^{1}$University of California, San Diego \quad $^{2}$University of Rochester
}

\begin{document}

\maketitle

\begin{abstract}
Text-conditioned symbolic music generation has attracted growing interest with the development of generative models. Yet two questions remain underexplored: what aspects of music the conditioning text should describe, and how the corresponding data should be collected and trained to improve controllability and text–music alignment. We address these challenges with TASTE, a taxonomy-aware framework for symbolic music generation with a main focus on piano pieces. First, we introduce a taxonomy of musical elements that specifies how a piece can be captured through textual description in a musically grounded way. The taxonomy comprises a set of metadata tags (genre, emotion, section plan, etc.) together with a free-form description of music imagery. This structure decomposes textual input to match the generation goal more coherently and precisely. Second, we build a data pipeline for collecting, annotating, and retrieving public symbolic piano resources, yielding a corpus of 432K MIDI files labeled at multiple levels of granularity. Finally, we design a staged training procedure that leverages this multi-granularity data across two model families: an autoregressive model and a non-autoregressive diffusion model. Both objective and subjective evaluations show that our models achieve superior musical-element control, text-description alignment, and overall music quality. We will release the full pipeline, all annotated datasets, and the model checkpoints.

\end{abstract}

\section{Introduction}
\label{sec:intro}

Symbolic music generation offers a structured and editable complement to audio generation, well suited to composition, arrangement, and human-in-the-loop creation. Driven by recent advances in generative models, a growing number of systems now generate symbolic music directly from textual description, letting users steer composition with language. Yet before asking how to generate music from text, we should ask a more basic question: \textbf{what makes a good description of music?} 

Existing text-conditioned systems seldom confront this question directly and tend toward one of two extremes. Some~\citep{text2midi,wu2026midillm} condition on free-form captions, which are expressive but whose musical content is uncontrolled and hard to quantify. Others~\citep{musecoco2023,notagen} restrict control to a fixed set of attributes such as key, tempo, instrument, and genre, which are precise but cover only a narrow portion of what a user wish to express.



These two extremes are symptoms of a deeper issue: a textual description of music conveys two kinds of information, and each design captures only one. The first is objective and well-defined musical attributes such as key, genre, emotion, and section plans of structure, which rest on a shared vocabulary and admit discrete labels. The second is subjective and open-ended music imagery that evokes a scene, a mood, or the symbolic impression conjured by music pieces. A real-world music description from users usually interleaves both. Consider ``\textit{an anxious, arpeggiated passage with flexible timing, evoking a gathering storm that breaks into release}'': it combines discrete attributes with an imagined scene and a temporal progression, reducing neither to a few metadata fields nor to an undifferentiated caption. We call the resulting mismatch between the descriptions current systems can handle and the intentions users actually hold \textbf{the description-intention gap}.

To close this gap, we should call for an explicit taxonomy of musical description that unifies both kinds of information while keeping them distinct. Such a taxonomy serves two purposes at once. For users, it offers a principled guide for expressing intent with sufficient scope and controllability. For models, it provides a structured training signal, so that generation will more faithfully reflect what a description specifies. Concretely, it separates a set of structured attribute tags from a free-form imagery description, allowing each to be annotated, learned, and evaluated on its own terms. Besides, we should also admit that sometimes imagery inevitably carries traces of some attributes, such as emotion, which is harmless as long as it also contains the correct attribute tags,



Answering the first question reshapes the second: \textbf{how can we collect data annotated in our designed taxonomy?} Existing datasets rarely provide such labels. MidiCaps~\citep{melechovsky2024midicaps} builds captions from metadata and extracted features, which favors properties easy to read off a MIDI file, leaving music imagery uncovered; large-scale MIDI collections like ~\citep{bradshaw2025ariamidi} offer abundant data but none of the taxonomy labels we need; and piano-specific labeled datasets~\citep{bang2024piast} remain small. Closing this gap calls not only for adding a few fields to an existing dataset, but also for a data pipeline organized around the taxonomy to collect public symbolic piano data, annotate each piece by matching each condition to suitable evidence, and propagate labels from reliably paired data to weakly labeled data at scale. This yields multiple granularities so that training will not leave uncertain supervision dominate.

To address these challenges by proposing \textbf{TASTE}, a taxonomy-aware framework that connects musical intent, data construction, and model training for symbolic piano generation. The taxonomy includes \textbf{tags} for objective, finite-vocabulary attributes, a \textbf{section plan} for temporal form, and an \textbf{imagery description} for the subjective scene or impression a piece evokes. The pipeline and training recipe answer how such data can be collected and used: we annotate each taxonomy channel from the evidence best suited to it, propagate labels from reliably paired data, and yield a corpus of 432K MIDI files labeled at multiple granularities. Our contributions are threefold:


\begin{enumerate}[leftmargin=7mm]
    \item We propose an intent-oriented taxonomy of music description that separates objective tags, section plans, and music imagery. We tie each channel to its annotation evidence and evaluation to narrow the description-intention gap.
    \item We build a taxonomy-driven data pipeline to match each condition to a suitable source of evidence and propagate labels across tiers, yielding a three-tier piano corpus of 10K, 51K, 371K pieces at multi-granular labels.
    \item We design a staged training recipe, instantiate it in both autoregressive and latent diffusion models. On both objective and subjective evaluations, TASTE achieves the superior performance on attribute agreement, full-prompt alignment, temporal structure, and musicality to other baselines. Music samples and other details can be found at \href{https://taste-piano.github.io/}{https://taste-piano.github.io/}.
\end{enumerate}

\section{Related Work}
\label{sec:related}

\paragraph{Text-conditioned Symbolic Generation.}
Unconditional piano models generate expressive performances without textual control~\citep{wu2023compose,bradshawscaling}. Attribute-conditioned generators support emotion~\citep{hung2021emopia,emo}, composer style~\citep{generalitymastery2025,compvector2026}, genre~\citep{lehmkuhl2025pianotransformer}, and composition identity~\citep{notagen}. FIGARO combines learned and expert features, while MuseCoco maps text to fixed attributes~\citep{figaro,musecoco2023}. MidiCaps and MetaScore construct captions from extracted features and community metadata, respectively~\citep{melechovsky2024midicaps,xu2025metascore}. Using MidiCaps, text2midi introduces caption cross-attention, while MIDI-LLM adapts a pretrained language model for text-to-MIDI generation~\citep{text2midi,wu2026midillm}.

\paragraph{Descriptions of music and piano performance.}
MusicCaps and Song Describer provide expert and listener-written captions~\citep{agostinelli2023musiclm,manco2023songdescriber}. Piano-focused work pairs performances with text, aligns audio, MIDI, and language for retrieval, and supports text-guided rendering~\citep{bang2024piast,bang2025pianobind,zhang2025renderbox}. TASTE instead organizes descriptions into separately supervised tags, section plans, and imagery for symbolic generation.

\paragraph{Weak supervision and multimodal distillation.}
Confident learning detects likely label errors~\citep{northcutt2021confident}, language-aligned audio and symbolic representations support probes and retrieval~\citep{wu2023laionclap,wu2025clamp3}, and section fusion identifies recurring sections~\citep{salamon2021deep}. TASTE coordinates these methods through its taxonomy, matching fields to suitable evidence and selectively revising earlier labels under restricted rules.

\section{Taxonomy of Music Description}
\label{sec:position}
\begin{figure}
    \centering
    \vspace{-3mm}
    \includegraphics[width=0.9\linewidth]{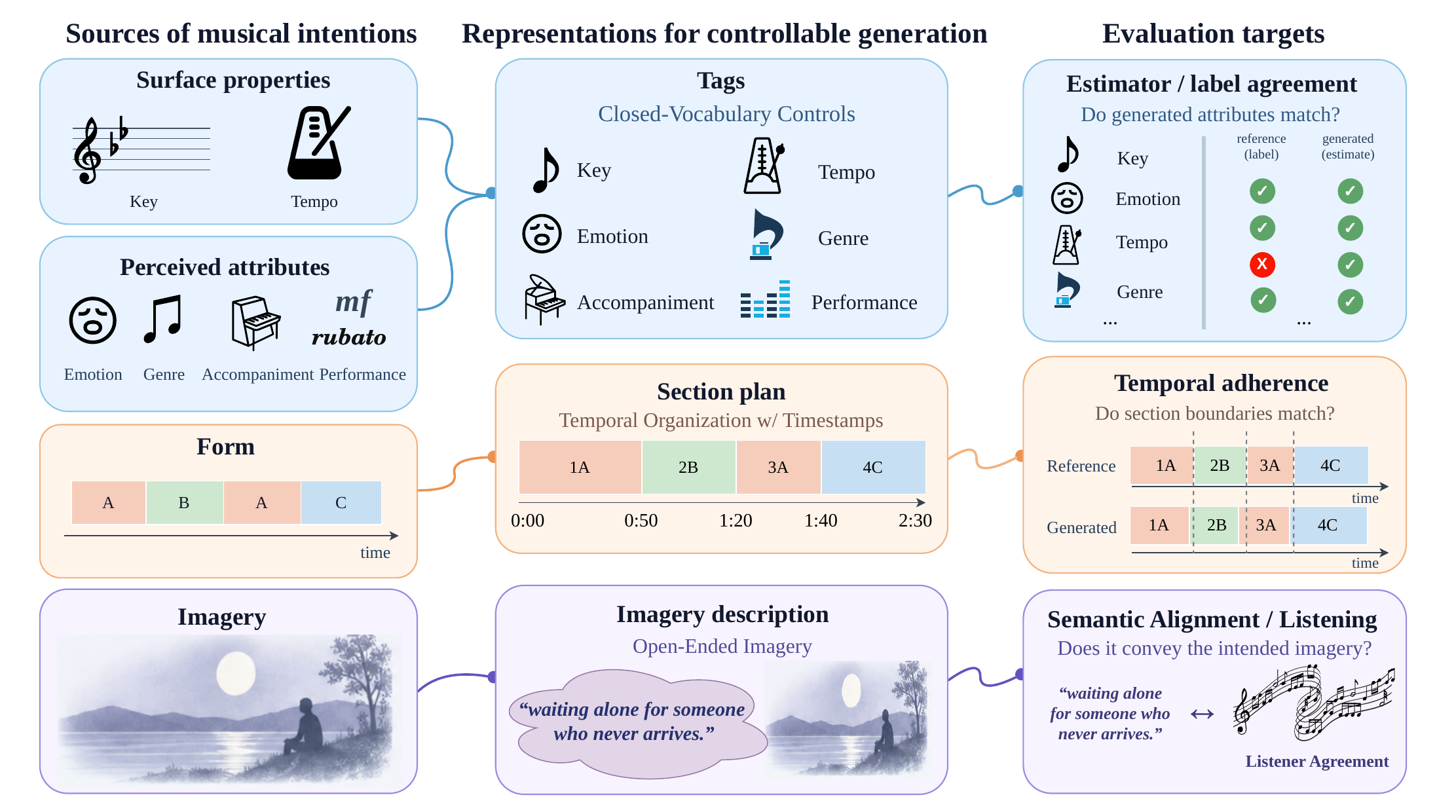}
    \vspace{-3mm}
    \caption{The TASTE taxonomy. Sources of musical intention (left) map to three condition channels (middle), each with its own evaluation target (right).}
    \label{fig:taxonomy}
    \vspace{-3mm}
\end{figure}

How a description is formatted has consequences on two sides: it bounds what a user can express, and determines what a model can be trained to control. A useful taxonomy should reflect how the aspects of a music description genuinely differ: in the vocabulary they draw on, in whether they hold globally or over time, and in the evidence their labels can rely on. These yield three channels from musical intentions in the left and middle of Figure~\ref{fig:taxonomy}: tags, section plan, and imagery description.

\subsection{From musical intentions to taxonomy channels}

\paragraph{Tags.}
Tags cover properties that fit a closed vocabulary and hold for the piano piece globally. They include surface properties namely key and tempo, and perceived attributes namely emotion, genre, accompaniment, and performance style. Among these properties, key and tempo are well handled by algorithmic estimation, while emotion or performance style requires perceptual interpretation. Performance style crosses two established dimensions of expressive performance~\citep{gabrielsson2003performance,widmer2004expressive} into four classes: steady, flexible, connected, or detached. Because each field has a predefined label space, extracted attributes can be compared field by field.

\paragraph{Section plan.}
Musical form is objective as one can verify from the notes whether a section returns. In addition to global tags, section plan represents as an ordered sequence of section identities with durations. Letters A--H identify sections within a piece, and a repeated letter marks recurrence. This ties form control to boundary and self-similarity analysis~\citep{foote2000novelty,salamon2021deep}, so that boundary placement and recurrence can be evaluated separately.

\paragraph{Imagery description.}
The remaining intentions resist any finite label set (except partially from emotion). Human beings may ask for music that conjures a scene, feeling, or symbolic impression from visual imagery~\citep{juslin2008mechanisms}. An imagery description expresses such that the music may evoke, e.g., ``waiting alone for someone who never arrives.'' It differs from tags in kind, not merely in vocabulary but also in its one-to-many mapping, as many descriptions may suit one piece and none is uniquely correct. Yet it is not unconstrained, for instance, a storm scene normally is open to anxious or angry emotions, but clearly not calm. The imagery channel is therefore a language condition whose target is perceptual compatibility rather than a single truth.

\subsection{Connecting representation, annotation, and evaluation}

The separation matters beyond the model interface, because each channel calls for different supervision and evaluation. Tags can be annotated and audited within finite vocabularies, section plans require temporal alignment, and imagery descriptions need contextual or perceptual grounding. Evaluation follows the same division instead of a single caption-similarity score (Figure~\ref{fig:taxonomy}, right): estimators and critics read back generated attributes, realized structure is compared with the plan, and listeners judge imagery, with text--music similarity as a joint proxy for tags and imagery.

\section{Multi-Tier Annotation Pipeline}
\label{sec:pipeline}

\begin{figure}[t]
    \centering
    \vspace{-3mm}
    \includegraphics[width=0.9\linewidth]{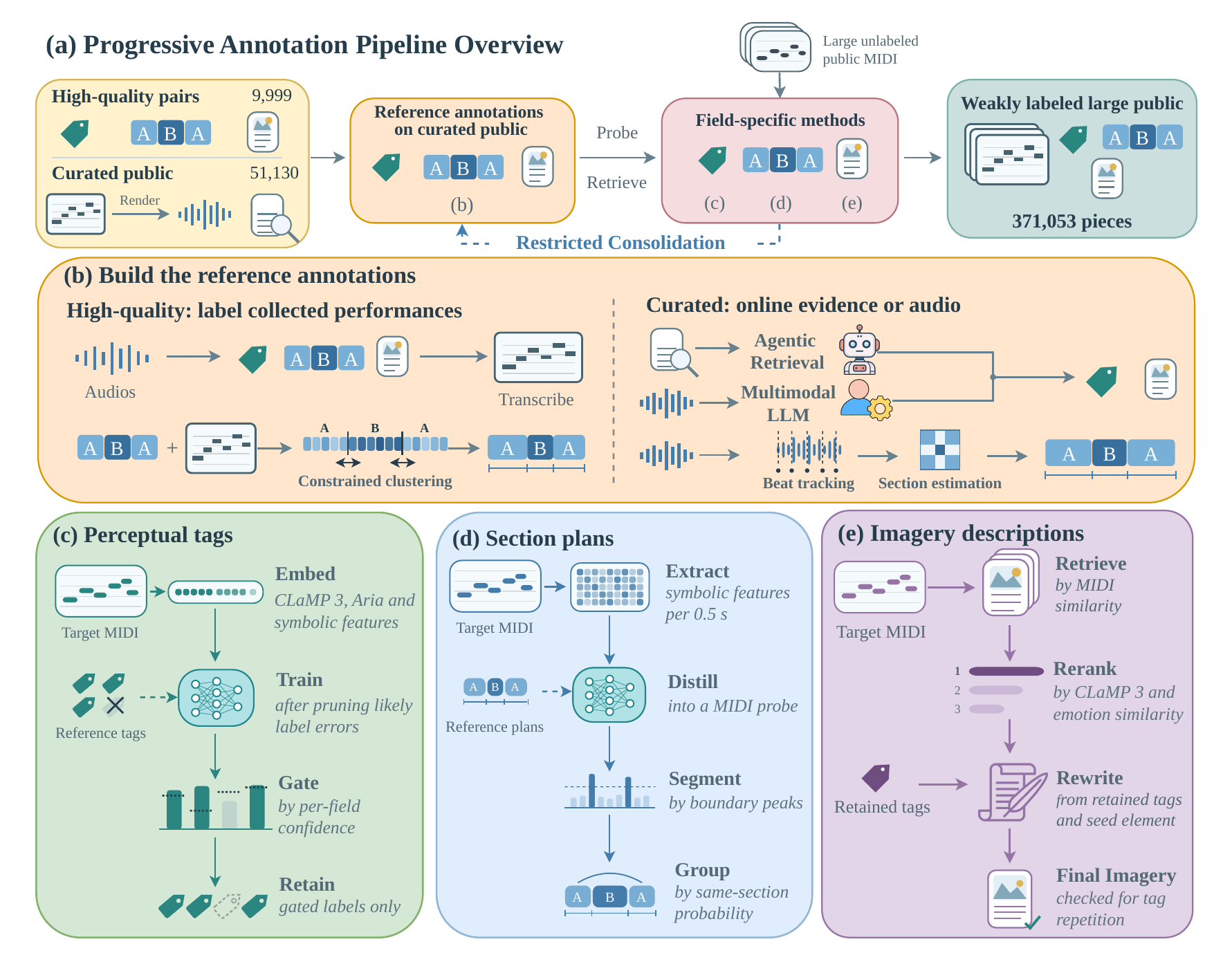}
    \vspace{-5mm}
    \caption{Progressive annotation across three tiers. Forward arrows gain scale at the cost of label reliability, while the dotted backward arrow is the restricted consolidation pass of Section~\ref{sec:pipeline:backward}.}
    \label{fig:pipeline}
    \vspace{-5mm}
\end{figure}

The annotation pipeline addresses how to collect data labeled in our taxonomy. We organizes 432K MIDI files into three tiers that trade label reliability for scale (Figure~\ref{fig:pipeline} and Table~\ref{tab:tiers}). We first build reference annotations in the high-quality and curated tiers by matching each taxonomy channel to suitable evidence, then propagate them to the large public tier through field-specific weak supervision, and finally revise selected earlier labels in a restricted backward pass.

\subsection{Reference annotations in the high-quality and curated tiers}
\label{sec:pipeline:curated}

We collect high-quality tier from public AI generated piano performances, and hire people to annotate four perceptual tags, imagery description and section sequence. We then transcribe the audio to MIDI with Transkun~\citep{transkun}. Of 9,999 collected pairs, 9,949 have an imagery description, and we hold out 100 of them as evaluation prompts.

In the curated tier, we collect multiple public symbolic music dataset in total of 51,130 pieces and infer their labels through one of two annotation routes. For 40,652 pieces, a language-model agent searches the web for evidence about the work using identity metadata such as composer, title, and movement, annotating each field the evidence supports and abstaining otherwise. The remaining 10,478 pieces lack identity metadata or receive at most one field from the agent. For them, a multimodal LLM describes the rendered audio, and a language model normalizes this description into tags and an imagery description. On both routes, imagery describes a plausible association without restating the tags, and a genre given by the source dataset takes priority (Appendix~\ref{app:pipeline}).

\begin{table}[t]
\centering
\begin{minipage}[t]{0.52\textwidth}
\centering\scriptsize
\caption{Annotation tiers and the label evidence.}
\label{tab:tiers}
\vspace{3mm}
\setlength{\tabcolsep}{3pt}
\begin{tabularx}{\linewidth}{@{}l r >{\raggedright\arraybackslash}X@{}}
\toprule
Tier & Pieces & Evidence source\\
\midrule
High-quality paired & 9,999 & Paired tags, imagery, section labels; estimated key, tempo, timing\\
Curated public & 51,130 & Web evidence or audio descriptions\\
Large public & 371,053 & Gated tag and section probes, imagery retrieval\\
\bottomrule
\end{tabularx}
\end{minipage}%
\hfill
\begin{minipage}[t]{0.44\textwidth}
\centering\scriptsize
\caption{Tag precision and tier coverage.}
\label{tab:large-tier-gates}
\vspace{2.2mm}
\setlength{\tabcolsep}{4pt}
\begin{tabular}{@{}lccc@{}}
\toprule
\textbf{Field} & \textbf{Precision} & \textbf{Coverage} & \textbf{Binding rule}\\
\midrule
Emotion & 0.600 & 86.6\% & precision \\
Genre & 0.936 & 100.0\% & precision \\
Accompaniment & 0.538 & 75.0\% & coverage floor \\
Performance & 0.588 & 75.0\% & coverage floor \\
\bottomrule
\end{tabular}
\end{minipage}
\end{table}

We then derive key, tempo, and timed section plans for both tiers. Key is estimated from MIDI pitch-class profiles~\citep{temperley2007music} and tempo from beats tracked in audio with Beat This!~\citep{foscarin2024beatthis} (Appendix~\ref{app:estimators}). A timed plan needs boundaries as well as identities that encode recurrence. High-quality pairs supply the section sequence but no timestamps, so we place boundaries by constrained clustering over symbolic and Aria features~\citep{bradshawscaling}, alternating dynamic-programming boundary placement with centroid re-estimation as in Lloyd's algorithm~\citep{lloyd1982least}. For curated performances, initial plans come from the tracked beats and deep-embedding section fusion~\citep{salamon2021deep}.

\subsection{Scaling annotations through field-specific weak supervision}
\label{sec:pipeline:scaling}

Running web search or audio annotation for each of the 371,053 pieces in the large tier would be costly. We instead propagate the reference annotations with a method suited to each channel: classification for perceptual tags, temporal prediction for section plans, and retrieval with rewriting for imagery descriptions. Key and tempo are estimated directly from MIDI (Appendix~\ref{app:estimators}).

\paragraph{Noise-aware tag probes.}
We predict perceptual tags with probes on CLaMP~3~\citep{wu2025clamp3} and Aria~\citep{bradshawscaling} embeddings and symbolic statistics of timing, dynamics, polyphony, and pedal use. Class-balanced linear classifiers predict emotion, genre, and accompaniment, and two binary classifiers for timing and articulation jointly determine performance style. We develop the probes on the 61,129 pieces of the two smaller tiers: we select features and calibrate per-field confidence thresholds on the validation set, prune likely label errors with a confident-learning rule on out-of-fold predictions~\citep{northcutt2021confident}, and refit the production probes on the pruned full set (Appendix~\ref{app:weak-supervision}). Where source metadata provide a genre, we map and keep it instead of the probe prediction.

A per-field confidence gate trades agreement with the reference annotations against coverage. For each field, we take the threshold that reaches 0.60 validation precision and relax it only as far as needed to label at least 75\% of the large tier. Predictions below the gate are omitted from the training conditions. Emotion and genre meet the precision target, whereas accompaniment and performance style are limited by the coverage floor and keep less precise labels (Table~\ref{tab:large-tier-gates}).

\paragraph{Section-plan distillation.}
We distill the timed section plans of the two smaller tiers into the \emph{structure probe}, a 1.8M-parameter transformer over frame-level symbolic features. A boundary head segments a performance, and a grouping head gives recurring segments the same section identity. This MIDI-native annotator needs neither audio nor metadata and thus covers the whole large tier.

\paragraph{Imagery retrieval and rewriting.}
For each target performance, we first retrieve candidate descriptions from the smaller tiers by Aria embedding similarity~\citep{bradshawscaling}. A ridge regression projects the target's Aria embedding into the CLaMP~3 text space~\citep{wu2025clamp3}, and we rerank the candidates by their similarity to this projection and by agreement between the two pieces' predicted emotion distributions. A language model then adapts the selected description to the target, keeping the mood and energy conveyed by the retrieved description and the target's predicted tags while changing the concrete setting (Appendix~\ref{app:imagery}).

\subsection{Restricted backward consolidation}
\label{sec:pipeline:backward}

The trained tag and structure probes can also revise the smaller tiers, filling gaps and correcting labels whose evidence is indirect, and we apply them conservatively. A gated prediction may replace any perceptual tag on audio-annotated pieces, but on evidence-annotated pieces, whose web evidence rarely covers texture or rendition, it may only fill or replace accompaniment and performance style, with a stricter threshold for replacing performance labels (Appendix~\ref{app:backward}). The structure probe replaces the curated tier's initial audio-derived plans, but on high-quality pairs it refines only the boundary timestamps and keeps the supplied identities and order. High-quality-tier tags and all imagery descriptions remain unchanged. The backward pass thus aligns labeling conventions across tiers while maintaining the reliability of each field.

\section{Models and Staged Training}
\label{sec:models}

\begin{figure}
    \centering
    \includegraphics[width=0.98\linewidth]{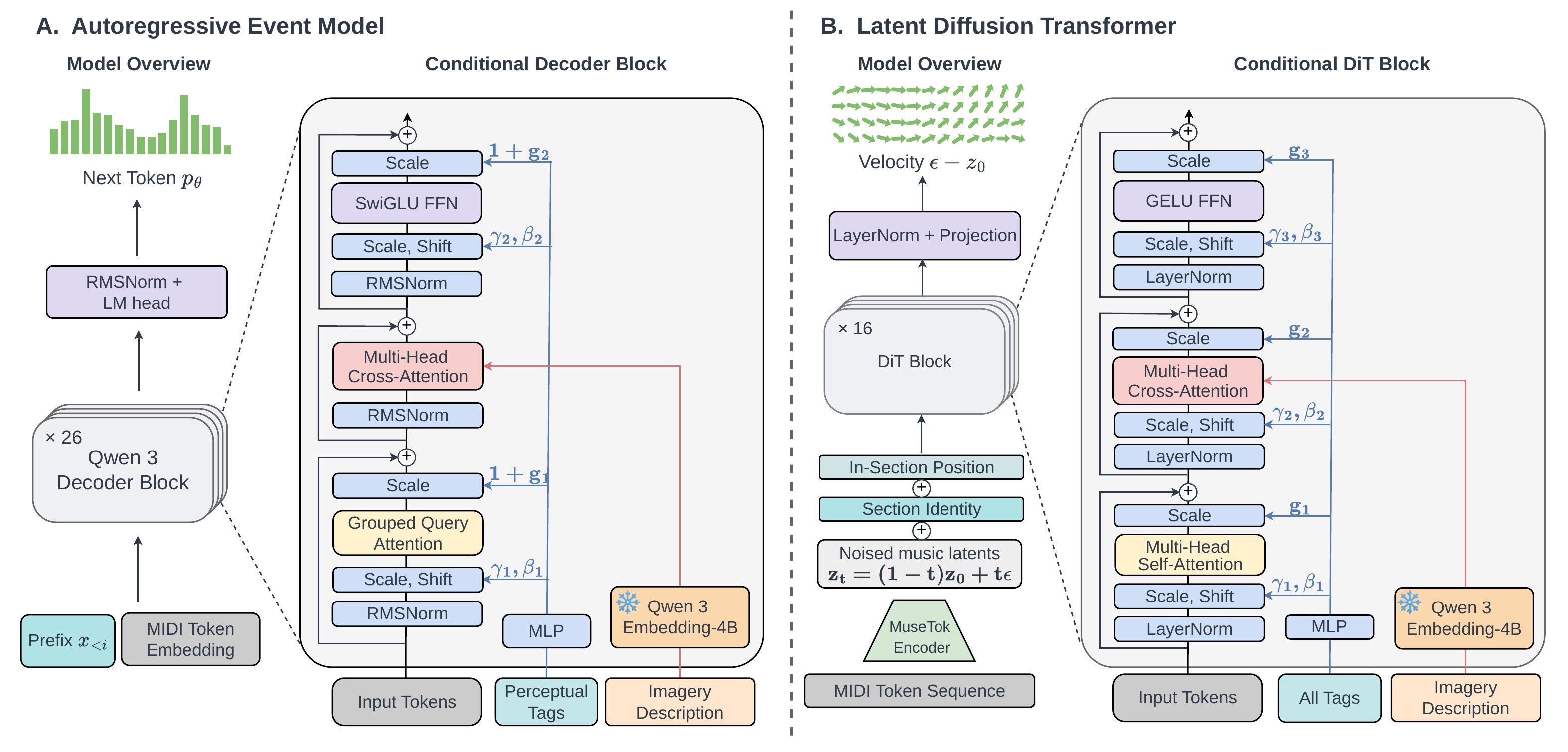}
    \vspace{-3mm}
    \caption{TASTE model families. (A) AR reads key, tempo, and the section plan as a token prefix. (B) DiT receives the section plan as per-position embeddings. In both, summed tag embeddings drive adaptive normalization, and imagery enters via cross-attention to a frozen text encoder.}
    \label{fig:architecture}
\end{figure}

The taxonomy specifies what information conditions generation, and the data tiers determine how supervision for these conditions enters training. We instantiate this design in two model families (Figure~\ref{fig:architecture}), an autoregressive decoder (AR) and a non-autoregressive latent diffusion transformer (DiT), which both keep the three channels separate while adapting each channel to the architecture.

\subsection{Piano representation}

Both models build on an Aria-style tokenizer~\citep{bradshawscaling} with a 25\,ms grid (Figure~\ref{fig:midi-tokenization} in Appendix~\ref{app:representation}). Each note is a combined pitch--velocity token with 12 velocity levels, followed by onset and duration tokens, and pedal events have their own tokens in both models. Window markers every five seconds anchor onsets in absolute time, replacing the bar tokens of metrical tokenizers~\citep{remi,cp}. In AR, key, tempo, and the section plan form an in-sequence prefix, in which each plan entry gives a section identity and its duration in windows. The active section identity is repeated after every window marker, so the model receives both a prospective plan and a local structural cue. The DiT's codec encodes only the note and pedal events, and the DiT receives key, tempo, and the plan as separate conditions (Section~\ref{sec:dit}).

\subsection{Autoregressive Model}
\label{sec:autoregressive}

The AR backbone is a 0.89B-parameter Qwen3-style decoder~\citep{yang2025qwen3} with a 4,096-token context. We pretrain it for five passes over about 817K performances of the pruned Aria-MIDI subset~\citep{bradshaw2025ariamidi}, with section-aligned crops and pitch and tempo augmentation, for 4.1M sequences and 11.44B tokens in total. Because key, tempo, and section plans can be estimated for any MIDI file (Section~\ref{sec:pipeline}), their tokens are present from pretraining, so the model learns how they relate to musical events before the remaining conditions are introduced.

The four perceptual tags inject to every decoder layer through AdaLN-Zero-style modulation~\citep{peebles2023dit}, with their embeddings summed and mapped to a 512-dimensional code. The imagery description enters via cross-attention to the token states of a frozen Qwen3-Embedding-4B~\citep{zhang2025qwen3embedding}. Adding these modules gives a 1.235B conditioanl generator (Appendix~\ref{app:models}).

\subsection{Non-autoregressive latent diffusion}
\label{sec:dit}

The DiT generates in a continuous latent space. A MuseTok-based window VAE~\citep{musetok} encodes each one-second window of a performance, with a 0.5\,s hop, into a 128-dimensional latent. A 16-layer, 397.6M-parameter diffusion transformer~\citep{peebles2023dit} is trained on these latents with a rectified-flow objective~\citep{liu2023rectifiedflow}: given a music latent $z_0$, Gaussian noise $\epsilon$, and a timestep $t$, it receives $z_t=(1-t)z_0+t\epsilon$ and predicts the velocity $\epsilon-z_0$. Like AR, the DiT is first pretrained on the same corpus with key, tempo, and section plans.

The DiT differs from AR in how key, tempo, and the section plan enter. All six tags, key and tempo included, are embedded, summed, and combined with the timestep embedding into the AdaLN-Zero code that modulates every block (Appendix~\ref{app:dit-conditioning}). The section plan is aligned with the latent sequence: each position receives embeddings of its section identity, position within the current section, and its absolute position~\citep{transformer}. The imagery description enters through cross-attention to the same states as in AR. At inference, we apply classifier-free guidance~\citep{ho2022classifierfree} and decode latents with a grammar-constrained MuseTok-based decoder (Appendix~\ref{app:models}).

\section{Evaluation}
\label{sec:results}

Following the division of Section~\ref{sec:position}, we first test whether each taxonomy channel is realized in the generated music, through agreement with the requested tags and descriptions and adherence to section plans. We then examine music quality and listener judgments. Appendix~\ref{app:stages} analyzes the effect of staged training and ablation expereiment with no large weak tier training (pre-training--curated tier--high-quality tier).

\subsection{Evaluation setup}

\paragraph{Generation protocol and baselines.}
We use the 100 held-out prompts of the high-quality tier and draw eight samples per prompt from each model. Every prompt contains six tags and an imagery description. AR generates its own section plan, while DiT additionally receives the plan and duration of the reference performance.

The conditional baselines are MuseCoco~\citep{musecoco2023}, text2midi~\citep{text2midi}, MIDI-LLM~\citep{wu2026midillm}, EMO-Disentangler~\citep{emo}, and Piano Transformer~\citep{lehmkuhl2025pianotransformer}, each receiving only the controls it supports. The caption-conditioned text2midi and MIDI-LLM receive all tags and the imagery in one caption, with non-piano tracks masked at decoding. Aria~\citep{bradshawscaling} and Compose \& Embellish~\citep{wu2023compose} serve as unconditional baselines for music quality. Appendix~\ref{app:eval} details the evaluation pipeline and data processing.

\paragraph{Condition agreement.}
Key and tempo are read back with the annotation estimators. For the perceptual tags, we train logistic-regression critics on CLaMP~3 and Aria embeddings of the 9,849 high-quality training pieces, and scoring the 100 reference performances with the same critics shows how reliably each field can be read back. Full-prompt alignment is the cosine similarity, under CLaMP~3~\citep{wu2025clamp3} and LAION-CLAP~\citep{wu2023laionclap}, between the generated music and a caption that combines the perceptual tags with the imagery description.

\paragraph{Music quality.}
We compute Fr\'echet Audio Distance in VGGish space~\citep{kilgour2019fad} and Fr\'echet Music Distance in CLaMP~3 space~\citep{retkowski2024fmd,wu2025clamp3} against MAESTRO~\citep{hawthorne2019maestro} for classical and POP909~\citep{wang2020pop909} for pop. Audiobox Aesthetics scores for production quality (PQ) and content enjoyment (CE)~\citep{tjandra2025audiobox} serve as proxies for listener ratings. All audio-based metrics, including those of the reference performances, use audio rendered from MIDI with shared synthesizer settings (Appendix~\ref{app:eval}).

\subsection{Controllability and text--music alignment}

\begin{table}[t]
\centering\small
\vspace{-3mm}
\caption{Tag agreement and full-prompt alignment ($\uparrow$). Reference: the 100 held-out high-quality MIDIs, excluded from ranking. $m$: mode-only key agreement. $q$: emotion agreement over four quadrants. $\dagger$: agreement with the key declared by the model. Bold and underline mark the best and second best.}
\vspace{3mm}
\label{tab:control}
\resizebox{\linewidth}{!}{
\begin{tabular}{@{}lcccccccc@{}}
\toprule
\textbf{Model} & \textbf{Key} & \textbf{Tempo} & \textbf{Emotion} & \textbf{Genre} & \textbf{Accomp.} & \textbf{Perf.} & \textbf{CLaMP~3} & \textbf{CLAP}\\
\midrule
Reference & 1.000 & 0.820 & 0.840 & 0.830 & 0.550 & 0.530 & 0.132 & 0.338\\
\midrule
MuseCoco & $0.582^m$ & 0.636 & $0.575^q$ & 0.386 & -- & -- & -- & --\\
text2midi & 0.181 & 0.456 & 0.276 & 0.380 & 0.205 & 0.270 & 0.054 & 0.273\\
MIDI-LLM & 0.340 & 0.713 & 0.285 & 0.370 & 0.201 & 0.226 & 0.059 & 0.292\\
EMO-Disentangler & $0.752^\dagger$ & -- & $\mathbf{0.611^q}$ & -- & -- & -- & -- & --\\
Piano Transformer & -- & -- & -- & 0.458 & -- & -- & -- & --\\
\midrule
\textbf{TASTE (AR)} & \underline{0.767} & \underline{0.730} & \underline{0.585} & \underline{0.509} & \underline{0.296} & \underline{0.273} & \textbf{0.141} & \underline{0.335}\\
\textbf{TASTE (DiT)} & \textbf{0.993} & \textbf{0.809} & 0.584 & \textbf{0.603} & \textbf{0.326} & \textbf{0.350} & \underline{0.123} & \textbf{0.347}\\
\bottomrule
\end{tabular}}
\vspace{-3mm}
\end{table}

Table~\ref{tab:control} shows that both TASTE models match or exceed the caption-conditioned baselines on all six tag fields and both full-prompt alignment scores. Emotion agreement roughly doubles, from at most 0.285 to 0.584--0.585, and CLaMP~3 alignment more than doubles. TASTE also exceeds the attribute-conditioned baselines on their own fields, except for EMO-Disentangler's quadrant-level emotion. Gains on both the discrete tags and the language-based alignment indicate that the two kinds of condition can be learned jointly within one model. The two families differ in emphasis: DiT, sampled with classifier-free guidance, leads on most tag fields, whereas AR, sampled without guidance, still attains the highest CLaMP~3 alignment and matches DiT on emotion.

Accompaniment and performance style remain the hardest fields. Even on the reference performances, the critics agree with their labels only slightly more than half of the time, and these are the two fields whose large-tier labels fall below the precision target (Table~\ref{tab:large-tier-gates}). Both observations point to label reliability as the limiting factor for these fields, in training as well as in evaluation.

\subsection{Temporal structure}

We evaluate section boundaries and recurrence with the structure probe of Section~\ref{sec:pipeline:scaling}, which predicts both boundaries and section identities, and a training-free Foote novelty curve~\citep{foote2000novelty}. \emph{Boundary F1} matches planned and detected boundaries one-to-one within 5\,s. \emph{Novelty contrast} divides the novelty near planned boundaries by the piece's median novelty. \emph{Pairwise F1} checks whether the plan and the detector assign the same section identity to each pair of frames, testing whether recurring segments are grouped together.

\begin{wraptable}[11]{r}{0.56\columnwidth}
\centering\small
\vspace{-3mm}
\caption{Temporal structure ($\uparrow$)}
\vspace{3mm}
\label{tab:structure}
\resizebox{\linewidth}{!}{%
\begin{tabular}{@{}lccc@{}}
\toprule
\textbf{System} & \textbf{Boundary F1} & \textbf{Contrast (random)} & \textbf{Pairwise F1}\\
\midrule
Reference & 0.896 & 2.52 (1.72) & 0.769\\
AR & 0.517 & 1.99 (1.69) & 0.608 \\
DiT & {0.868} & {3.18} (1.78) & {0.775}\\
\bottomrule
\end{tabular}
}
\vspace{-3mm}
\end{wraptable}

Table~\ref{tab:structure} shows that DiT closely realizes its supplied section plan. Its boundary F1 of 0.868 approaches the 0.896 of the reference performances, and its pairwise F1 matches the reference. Its novelty contrast of 3.18, against 1.78 at random positions and 2.52 for the references, shows clear changes in musical features at planned transitions. AR also places changes near its self-generated boundaries, but its smaller contrast gap (1.99 versus 1.69) and boundary F1 of 0.517 indicate a weaker link between the plan prefix and the events that follow. These results suggest that aligning the plan with the generated sequence supports reasonable realization of form, even for AR sampled without guidance (Appendix~\ref{app:structure-eval}).

\subsection{Music quality}

\begin{table}[t]
\centering\small
\vspace{-3mm}
\caption{Objective music quality. V/C: Fr\'echet distance in VGGish/CLaMP~3 space. Bold and underline mark the best and second best, excluding the reference row.}
\vspace{3mm}
\label{tab:quality}
\resizebox{\linewidth}{!}{
\begin{tabular}{@{}lcccccc@{}}
\toprule
\textbf{Model} & \textbf{MAESTRO V $\mathbf{\downarrow}$} & \textbf{MAESTRO C} $\downarrow$ & \textbf{POP909 V }$\downarrow$ & \textbf{POP909 C} $\downarrow$ & \textbf{PQ} $\uparrow$ & \textbf{CE} $\uparrow$\\
\midrule
Reference & 0.497 & 193.427 & 0.662 & 239.368 & 8.007 & 7.333\\
\midrule
MuseCoco & 1.045 & 372.282 & 0.964 & 280.778 & 7.709 & 6.879\\
text2midi & 1.067 & 296.012 & 1.113 & 277.089 & 7.789 & 6.953\\
MIDI-LLM & 1.425 & 282.214 & 2.025 & 250.544 & 7.586 & 6.840\\
EMO-Disentangler & 0.979 & 356.460 & \textbf{0.560} & 250.884 & 7.944 & \underline{7.448}\\
Piano Transformer & 0.745 & \underline{196.619} & 1.542 & 247.119 & 7.761 & 7.180\\
Aria (unconditional) & 0.831 & 320.512 & 1.114 & 241.098 & 7.864 & 6.868\\
Compose \& Embellish & 2.049 & 383.140 & 0.898 & 219.129 & \textbf{7.984} & \textbf{7.502}\\
\midrule
\textbf{TASTE (AR) }& \underline{0.426} & \textbf{170.697} & \underline{0.605} & \underline{201.652} & \underline{7.946} & 7.410\\
\textbf{TASTE (DiT)} & \textbf{0.309} & 239.704 & 0.808 & \textbf{196.491} & 7.945 & 7.319\\
\bottomrule
\end{tabular}}
\vspace{-3mm}
\end{table}

Table~\ref{tab:quality} shows that TASTE combines broad condition control with competitive music quality. Both models rank first or second in MAESTRO audio distance and POP909 symbolic distance, and both improve on the unconditional Aria model in all six metrics. AR has the lowest MAESTRO symbolic distance and DiT the lowest MAESTRO audio distance, suggesting that the two representations preserve different aspects of the reference distributions. Their production-quality scores are nearly identical (7.946 and 7.945) and close to the best, Compose \& Embellish (7.984). Content enjoyment favors Compose \& Embellish and EMO-Disentangler, but our models remain competitive.

\subsection{Listening study}
\label{sec:listening}

A listening study measures perceived quality and imagery alignment for both TASTE models, text2midi, MIDI-LLM, and the reference performances, using 24 held-out prompts and the first sample of each system. Twenty-two participants each judged 15 random blind 30-second excerpts, rating overall quality from 1 to 5 and then choosing which of two imagery descriptions fits better.

\begin{wraptable}{r}{0.5\columnwidth}
\vspace{-7mm}
\centering\small
\caption{Listening study.}
\vspace{2mm}
\label{tab:listening}
\resizebox{\linewidth}{!}{%
\begin{tabular}{@{}lcc@{}}
\toprule
\textbf{System} & \textbf{Quality (1--5)} $\uparrow$ & \textbf{Imagery accuracy (\%)} $\uparrow$\\
\midrule
Reference & 3.85 $\pm$ 0.85 & 69.7\\
\midrule
text2midi & 2.71 $\pm$ 0.91 & 54.5\\
MIDI-LLM & \underline{2.82 $\pm$ 1.08} & 51.5\\
\midrule
TASTE (AR) & \textbf{3.26 $\pm$ 0.79} & \textbf{80.3}\\
TASTE (DiT) & 2.73 $\pm$ 1.00 & \underline{75.8}\\
\bottomrule
\end{tabular}
}
\end{wraptable}

Table~\ref{tab:listening} shows that listeners match TASTE outputs to their imagery descriptions even more often than the reference. By contrast, text2midi and MIDI-LLM stay near chance, in line with their low emotion agreement (Table~\ref{tab:control}). Even for the reference performances, listeners choose the unaligned description in 30\% of trials, which further supports that imagery admits multiple valid descriptions (Section~\ref{sec:position}). Quality ratings separate the two TASTE models more than the objective metrics do (Table~\ref{tab:quality}). AR is rated highest among all models, whereas DiT is rated on par with the worst baseline although it outperforms that baseline on all automatic metrics. DiT thus aligns with the imagery about as reliably as AR, but with lower perceived quality.

\section{Conclusion}
\label{sec:conclusion}

We presented TASTE, which addresses the description-intention gap by asking what makes a good description of music and how data for such descriptions can be collected. Its taxonomy separates tags, section plans, and imagery descriptions, a multi-tier pipeline annotates each channel from suitable evidence, and staged training carries these labels into autoregressive and diffusion models. With the channels kept separate from annotation to generation, both models follow tags and imagery more closely than caption-conditioned baselines, listeners match their outputs to the intended imagery well above chance, and AR's music quality remains competitive.

Yet the results suggest that accompaniment and performance style, the fields with the least reliable labels, remain the hardest to control, making label reliability the next bottleneck. These findings motivate treating a description as a structured specification whose channels each receive suitable evidence and model pathway. TASTE provides a basis for extending this specification toward richer musical intentions while preserving the freedom to realize the same request in different ways.

\section*{AI Use Statement}
We used large language models (LLMs) to edit and polish the writing of this paper. All research decisions, including the method design, experiments, and conclusions, were made by the authors, who verified every result and all text. The LLMs that annotate data in our pipeline (Section~\ref{sec:pipeline}) are part of the proposed method and are described in Appendix~\ref{app:pipeline}.

\section*{Ethics Statement}
This work generates symbolic piano music from descriptive conditions. Our corpora combine publicly released piano datasets with high-quality paired data, and the released annotations will follow the licenses of their source datasets. No condition field encodes composer or performer identity, so the models are not designed to imitate specific artists. The listening study collected anonymous responses and recorded no personal information beyond self-reported years of musical training and the playback device. We are not aware of other ethical concerns specific to this work.

\section*{Reproducibility Statement}
We will release the codebase for annotation, corpus construction, training, inference, and evaluation, together with the annotated datasets and model checkpoints. Section~\ref{sec:pipeline} and Appendices~\ref{app:taxonomy}--\ref{app:pipeline} specify the label vocabularies, annotation routes, estimators, probe training, confidence gates, and retrieval. Section~\ref{sec:models} and Appendix~\ref{app:models} give the architectures, per-stage training configurations, and sampling settings. Section~\ref{sec:results} and Appendix~\ref{app:eval} describe the evaluation protocol, critics, baseline interfaces, rendering, and listening study. Data splits, listening prompts, and random-position baselines follow deterministic rules or fixed seeds. Because web-search and LLM-based annotation can vary with time, model version and random seed, the released annotations allow the training and evaluation stages to be reproduced without repeating these steps.

\bibliography{iclr2027_conference}
\bibliographystyle{iclr2027_conference}

\appendix
\section{Condition specification}
\label{app:taxonomy}

\subsection{Tag vocabularies and annotation scope}

Table~\ref{tab:vocabulary} defines the classes of the six tag fields. Each field takes one class, and an abstention becomes the null. On the audio route, the normalizer also applies three scope rules: emotion follows the affect of greatest cumulative duration, accompaniment and performance style describe the first clearly identifiable section, and explicitly broken chords take precedence over the broader ostinato class.

\begin{table}[htbp]
\caption{Tag vocabularies and annotation scope.}
\vspace{3mm}
\centering\small
\setlength{\tabcolsep}{4pt}
\begin{tabular}{@{}lc@{}}
\toprule
{\bfseries Field} & {\bfseries Classes and operational definition}\\
\midrule
Key & \parbox[t]{0.76\linewidth}{\raggedright 24 classes, 12 major and 12 minor tonics: one global estimate per piece.}\\
\addlinespace[1pt]
Tempo & \parbox[t]{0.76\linewidth}{\raggedright 3 classes from the estimated bpm: \emph{slow} $< 75$, \emph{moderate} $\in [75,115)$\,, \emph{fast} $\geq 115$.}\\
\addlinespace[1pt]
Emotion & \parbox[t]{0.76\linewidth}{\raggedright \emph{happy} brightness, playfulness, or triumph. \emph{sad} sorrow, longing, or grief. \emph{calm} serenity or reflection without persistent distress. \emph{angry} confrontation or defiance. \emph{anxious} tension, restlessness, or foreboding.}\\
\addlinespace[1pt]
Genre & \parbox[t]{0.76\linewidth}{\raggedright \emph{classical} Baroque through modern and neoclassical. \emph{jazz} ragtime, stride, swing, blues, jazz ballads. \emph{pop} pop-song covers, pop ballads, New Age and easy-listening styles. Hybrids take the strongest overall identity.}\\
\addlinespace[1pt]
Accompaniment & \parbox[t]{0.76\linewidth}{\raggedright \emph{chordal} block or voiced chords. \emph{ostinato} a repeated driving cell. \emph{arpeggiated} broken or rolling chords. \emph{sustained} held tones, pedal points, or sparse long harmony. \emph{contrapuntal} independently moving voices.}\\
\addlinespace[1pt]
Performance & \parbox[t]{0.76\linewidth}{\raggedright Two binary axes crossed into four classes: timing \emph{steady} (strict, even, motoric) versus \emph{flexible} (rubato, elastic, breathing), and articulation \emph{connected} (legato, singing, smooth) versus \emph{detached} (staccato, crisp, percussive).}\\
\bottomrule
\end{tabular}

\label{tab:vocabulary}
\end{table}

\subsection{Section identities and imagery}

Section identities are letters assigned in order of first appearance within each piece, so a repeated letter denotes recurring material. A piece uses at most eight identities (A--H), and each section's duration is counted in five-second windows.

Imagery descriptions express a scene, atmosphere, or experience without restating musical tags. On the evidence route, the scene is grounded in the retrieved context, and on the audio route, it is derived from the audio description. In the large tier, rewriting keeps the mood and energy conveyed by a retrieved description and the target's predicted tags while changing the setting (Appendix~\ref{app:imagery}).
\section{Corpus construction and validation}
\label{app:pipeline}

\subsection{Corpus accounting and splits}

Table~\ref{tab:source-composition} lists the sources of the curated tier. PIAST-YT and Pop2Piano recordings are transcribed to piano MIDI, while the other sources supply MIDI directly.

\begin{table}[htbp]
\caption{Source composition of the curated tier.}
\vspace{3mm}

\label{tab:source-composition}
\centering\footnotesize
\begin{tabular}{@{}lc@{}}
\toprule
{\bfseries Source} & {\bfseries Pieces}\\
\midrule
Aria-MIDI named subset~\citep{bradshaw2025ariamidi} & 32,522\\
PIAST-YT~\citep{bang2024piast} & 8,669\\
Pop2Piano cover index~\citep{choi2022pop2piano} & 5,866\\
PiJAMA~\citep{edwards2023pijama} & 2,777\\
POP909~\citep{wang2020pop909} & 909\\
EMOPIA-derived pieces~\citep{hung2021emopia} & 387\\
\midrule
Total & 51,130\\
\bottomrule
\end{tabular}
\end{table}

The large tier contains 371,053 pieces from the deduplicated Aria-MIDI release~\citep{bradshaw2025ariamidi}, and pretraining uses its pruned release. A note is that tiers are not disjoint by work.

The high-quality tier contributes 9,849 audio--text training pairs after removing 50 pairs with empty imagery and holding out 100 pairs for evaluation.

Probe selection and calibration use a 61,129-piece benchmark that joins the curated tier with all high-quality pairs and groups them by work identity. Clips of one recording share a group, and each high-quality piece forms its own group. These groups define a deterministic 80/20 split as well as the folds of five-fold pruning.

\subsection{Curated annotation protocols}
\label{app:annotation-protocols}

\paragraph{Evidence route.}
A Qwen3.6-35B-A3B agent issues at most two web searches and reads at most one page per piece, using only sources whose identity matches the piece: creator, catalogue number, and movement for classical works, and title and artist for songs. Typical sources are encyclopedia entries, program notes, and lyrics pages. It annotates each perceptual field separately and abstains when the evidence is insufficient. Source-provided genre and single-valued emotion labels take precedence. The imagery is written as an original scene grounded in the retrieved context.

\paragraph{Audio route and normalization.}
Pieces without identifiable work metadata go directly to audio annotation, as do pieces for which the evidence route accepts at most one field. Gemini 3.5 Flash describes the genre, affect, texture, timing, articulation, and development of the rendered audio. Qwen3.6-35B-A3B then normalizes this description into emotion, accompaniment, and performance style, following the scope rules of Appendix~\ref{app:taxonomy}, and into a short imagery description that keeps the scene or atmosphere but omits musical terminology. Genre takes the label of the source dataset.

\subsection{Key and tempo estimation}
\label{app:estimators}

Key is the major or minor key whose Kostka--Payne profile, as reported by \citet[Table~6.1]{temperley2007music}, correlates best with the piece's duration-weighted pitch-class profile. Durations are measured in seconds under the MIDI tempo map, percussion is excluded, and both profiles are centered and normalized before comparison.

In the smaller tiers, tempo is $60/\mathrm{median}(\Delta t)$ over the inter-beat intervals from Beat This!~\citep{foscarin2024beatthis} and tracks that fail or fall outside 20--400\,bpm use the MIDI estimator instead. This estimator, which also serves the whole large tier, is a simplified inter-onset-interval histogram inspired by \citet{dixon2001tempo}. It collects lag-$k$ onset differences for $k\leq8$ within 0.12--3.0\,s, folds their tempos into 50--180\,bpm, and smooths a 1\,bpm histogram weighted by $1/\sqrt{k}$. The estimate is the weighted mean within 2\,bpm of the peak.

\subsection{Tag probes and confidence gates}
\label{app:weak-supervision}

Each probe uses CLaMP~3 music embeddings (768 dimensions)~\citep{wu2025clamp3}, Aria embeddings (512)~\citep{bradshawscaling}, and 32 symbolic statistics summarizing timing, pitch, dynamics, polyphony, and pedal. We apply median imputation and standardization before class-balanced logistic regression. Performance style combines two binary probes, predicting flexible timing at $p_{\mathrm{flex}}\geq0.45$ and detached articulation at $p_{\mathrm{det}}\geq0.75$, with joint confidence
\[
\min\{\max(p_{\mathrm{flex}},1-p_{\mathrm{flex}}),\max(p_{\mathrm{det}},1-p_{\mathrm{det}})\}.
\]

Pruning follows a confident-learning rule~\citep{northcutt2021confident}. For supplied label $y_i$ and work-grouped out-of-fold probabilities $p_i(c)$, define the class self-confidence $\mu_c=|I_c|^{-1}\sum_{i\in I_c}p_i(c)$, where $I_c=\{i:y_i=c\}$. Example $i$ is removed only when $\arg\max_c p_i(c)\neq y_i$ and $p_i(y_i)<\mu_{y_i}$. Table~\ref{tab:pruning-ablation} compares the pruned and unpruned fits: pruning helps emotion, accompaniment, and articulation most, leaves genre and timing nearly unchanged, and slightly lowers the combined performance style. The production probes are refit on the retained labels of the full benchmark, and calibration uses train-side fits evaluated on the held-out split.

\begin{table}[htbp]
\caption{Tag-probe macro-F1 before and after pruning. Kept: fraction of training labels retained.}
\vspace{3mm}

\label{tab:pruning-ablation}
\centering\footnotesize
\resizebox{\linewidth}{!}{
\begin{tabular}{@{}lcccccc@{}}
\toprule
 & {\bfseries Emotion} & {\bfseries Genre} & {\bfseries Accomp.} & {\bfseries Timing} & {\bfseries Articulation} & {\bfseries Performance}\\
\midrule
Unpruned & 0.5353 & {0.9086} & {0.4445} & {0.6527} & {0.7256} & {0.5119}\\
Pruned & {0.5629} & {0.9130} & {0.4621} & {0.6523} & 0.7405 & {0.5027}\\
Kept & 55.3\% & 93.1\% & 50.8\% & 65.4\% & 81.0\% & -- \\
\bottomrule
\end{tabular}}

\end{table}

For field $f$, $\theta_f^{\mathrm{prec}}$ is the lowest validation confidence at which precision reaches 0.60 with at least 50 retained examples, and $\theta_f^{\mathrm{floor}}$ is the quantile of production scores that retains 75\% of the large tier. The applied gate is
\[
\theta_f=\min(\theta_f^{\mathrm{prec}},\theta_f^{\mathrm{floor}}),
\]
and predictions below it are omitted from the conditions. For accompaniment and performance style, the coverage floor binds and precision falls below the target (Table~\ref{tab:gate-details}). Genre uses mapped source metadata where available (276,948 pieces) and the probe elsewhere.

\begin{table}[htbp]
\caption{Confidence thresholds, validation precision, and large-tier coverage. Genre coverage includes source metadata.}
\vspace{3mm}

\label{tab:gate-details}
\centering\footnotesize
\setlength{\tabcolsep}{5pt}
\begin{tabular}{@{}lcccccc@{}}
\toprule
{\bfseries Field} & {\bfseries $\theta^{\mathrm{prec}}$} & {\bfseries $\theta^{\mathrm{floor}}$} & {\bfseries Applied} & {\bfseries Val.\ prec.} & {\bfseries Coverage} & {\bfseries Binding rule}\\
\midrule
Emotion & 0.7319 & 0.8807 & 0.7319 & 0.600 & 86.6\% & precision\\
Genre & 0.3968 & 1.0000 & 0.3968 & 0.936 & 100.0\% & precision\\
Accompaniment & 0.9873 & 0.8097 & 0.8097 & 0.538 & 75.0\% & coverage floor\\
Performance & 0.9319 & 0.9083 & 0.9083 & 0.588 & 75.0\% & coverage floor\\
\bottomrule
\end{tabular}
\end{table}

\subsection{Section supervision}
\label{app:section-supervision}

The audio teacher for the curated tier tracks beats with Beat This!~\citep{foscarin2024beatthis} and then applies beat-synchronous deep, few-shot, and CQT features with spectral clustering and section fusion~\citep{salamon2021deep}. We use hierarchy level 4, a 12-second minimum section, $\mu=\gamma=0.5$, and seed 0.

For high-quality pieces, the supplied letter sequence constrains segmentation. At a 0.5-second frame rate, segments sharing a letter share a centroid. With segment lengths $d_i$, letters $L_i$, interior cuts $b_i$, and total length $T$, the objective is
\[
\sum_{i=1}^{N}\sum_{t\in\mathrm{seg}_i}\|x_t-\mu_{L_i}\|^2
+\lambda_{\mathrm{dur}}\sum_{i=1}^{N}\frac{(d_i-T/N)^2}{T/N}
-\lambda_{\mathrm{nov}}\sum_{i=1}^{N-1}\mathrm{nov}(b_i).
\]
Dynamic programming optimizes the cuts for fixed centroids, and up to twelve Lloyd iterations~\citep{lloyd1982least} alternate between cut placement and centroid updates until convergence. We use $\lambda_{\mathrm{dur}}=0.10$, a 10-second novelty window, and an 8-second minimum segment. The first pass uses chroma, weighted by two, together with six texture statistics and $\lambda_{\mathrm{nov}}=0$. A second pass, initialized from those cuts, uses standardized, smoothed, unit-normalized Aria layer-7 features with $\lambda_{\mathrm{nov}}=1.0$. On 50 audited pieces with 236 reference boundaries, the mean absolute boundary error is 3.11\,s, compared with 6.3\,s for equal spacing.

The MIDI-native structure probe has 1,817,847 parameters: 26 symbolic features at 0.5-second resolution feed a four-layer, four-head transformer of width 192, with a boundary head and a 64-dimensional grouping head. Gaussian boundary targets use $\sigma=1.5$\,s. The grouping head reads the encoder states through a stop-gradient and learns a balanced same-section classification over teacher spans. Training uses 10 epochs, batch size 16, learning rate $3\times10^{-4}$, a 95/5 piece split, and threefold repetition of high-quality training pieces.

With a 5-second tolerance, a 12-second peak spacing, and each subset's validation-optimal threshold, the checkpoint that labeled the curated tier and the high-quality boundaries agrees with its teachers at a boundary F1 of 0.525 on 2,553 curated pieces and 0.878 on 503 high-quality pieces. It also labels the pretraining corpus and evaluation samples. Production labeling of the public tiers uses a 20-second spacing, boundary threshold 0.15, and agglomerative merging at same-section probability 0.65, capped at eight identities. Adjacent equal identities are merged, and a final span shorter than five seconds is absorbed. For high-quality pieces, each supplied boundary moves to the nearest unused probe peak (threshold 0.30, 12-second spacing) within $\pm4$\,s, provided that neighboring boundaries stay at least 12\,s apart and the boundary stays at least 5\,s from the end.

\subsection{Imagery retrieval and rewriting}
\label{app:imagery}

The candidate pool contains 60,783 descriptions with paired Aria and CLaMP~3 text embeddings. Ridge regression with regularization 1.0 maps normalized Aria vectors into the text space. For a query piece $q$, we retrieve its 50 nearest candidates in Aria space and select
\[
 j^*=\arg\max_{j\in\mathcal{N}_{50}(q)}
 \left[0.6\cos(Wa_q,t_j)+0.3\cos(e_q,e_j)\right],
\]
where $a_q$ is the query's Aria embedding, $t_j$ the CLaMP~3 text embedding of candidate $j$, and $e$ the predicted emotion distribution. Qwen3.5-27B-FP8 rewrites the selected scene, keeping the mood and energy conveyed by the scene and the target's predicted tags, including tempo, while changing the setting. Lexical checks reject rewrites that use musical terminology, drift into another script, or contain fewer than eight words, and such rewrites are regenerated. We also use 100 held-out prompt-MIDI pairs for evaluation.

\subsection{Restricted backward consolidation}
\label{app:backward}

On audio-annotated pieces, a prediction that passes the production gate of Table~\ref{tab:gate-details} may replace any of the four perceptual tags. On evidence-annotated pieces, the probes fill missing accompaniment and performance labels under the same gates and leave emotion and genre untouched. Existing accompaniment labels there are replaced at the production gate. Replacing an evidence-derived performance label requires a confidence of at least 0.9982, which limits such overrides to 3,013 of the 40,652 evidence-annotated pieces (7.41\%). Predictions below the applicable threshold leave the original label unchanged.

The structure probe supplies the curated plans and snaps high-quality boundaries with the production settings of Appendix~\ref{app:section-supervision}, preserving the supplied identities. High-quality tags and all imagery descriptions keep their original annotations.
\section{Model, training, and inference specification}
\label{app:models}

\subsection{Representation and augmentation}
\label{app:representation}

\begin{figure}[htbp]
    \centering
    \includegraphics[width=1\linewidth]{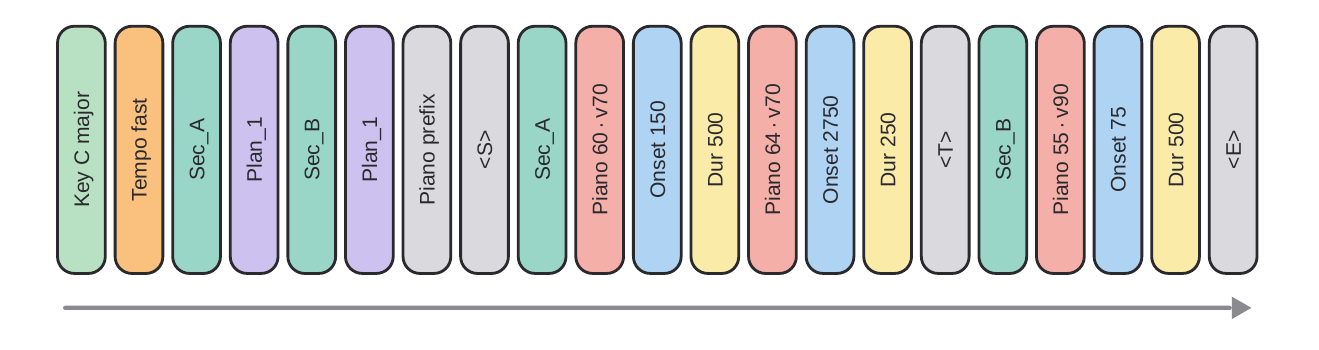}
    \vspace{-10mm}
    \caption{Tokenization example. The prefix holds key, tempo, and the section plan, where Sec\_A, Plan\_1 means that section A lasts one five-second window. <T> anchors each window and is followed by the active section identity, and <S> and <E> mark the start and end of the performance.}
    \label{fig:midi-tokenization}
\end{figure}

Figure~\ref{fig:midi-tokenization} shows a tokenized AR sequence. The vocabulary contains 2,120 active identifiers, padded to 2,176 embedding entries. Alongside note and timing events, it includes eight section identities, 24 key classes, three tempo classes, ten plan digits, and an interior-crop start marker. Prefix and event tokens share the 4,096-token context.

Pretraining crops start at a randomly selected eligible section and extend over whole sections. An overlong section is cut at a window boundary, and the plan is recomputed to fit. The end token is kept only when the crop reaches the end of the piece. Conditional rows always start at the true opening.

Augmentation draws an integer transposition uniformly from $[-5,+5]$ semitones and a time factor $a$ uniformly from $[0.90,1.10]$. Notes and key transpose together, event times and section boundaries scale by $a$, and a tempo of $b$\,bpm becomes $b/a$. Velocity is unchanged. AR pretraining uses one clean pass followed by four augmented passes. Every other stage, including DiT pretraining, places the clean pass last.

\subsection{Autoregressive conditioning}
\label{app:ar-conditioning}

Table~\ref{tab:ar-arch} specifies the Qwen3-style backbone~\citep{yang2025qwen3}. For each of the four perceptual fields $f$, the selected tag $a_f$ indexes an embedding table $E_f$ that includes a null entry for a missing or dropped label. The shared code is
\[
 c=\operatorname{SiLU}\!\left(W_c\sum_f E_f[a_f]+b_c\right).
\]
Each layer projects $c$ into a shift, scale, and gate $(\beta,\gamma,g)$ for its self-attention and feed-forward branches. For branch $F$ and pretrained normalization $N$, the update is
\[
 h'=h+(1+g)\odot F\!\left((1+\gamma)\odot N(h)+\beta\right).
\]
The modulation heads start at zero, so the $1+g$ gate preserves the pretrained branch at initialization.

Imagery enters through a cross-attention branch placed between self-attention and the feed-forward branch:
\[
 h'=h+W_o\operatorname{Attn}\!\left(W_q\operatorname{RMSNorm}(h),W_kS,W_vS\right).
\]
Here $S$ contains the final-layer token states of a frozen Qwen3-Embedding-4B~\citep{zhang2025qwen3embedding}, truncated at 64 tokens without pooling. A learned null context token is always prepended, and a missing imagery description attends to it alone. The output projection $W_o$ starts at zero, so adding both conditioning pathways initially preserves the pretrained model.

\begin{table}[htbp]
\caption{Autoregressive architecture. Parameter counts exclude the frozen text encoder.}
\vspace{3mm}

\label{tab:ar-arch}
\centering\small
\begin{tabular}{@{}lc@{}}
\toprule
{\bfseries Parameter} & {\bfseries Value}\\
\midrule
Layers / hidden width / context & 26 / 1,536 / 4,096\\
Query heads / key--value heads / head width & 24 / 6 / 64\\
Feed-forward width (SwiGLU) / embedding table & 6,144 / 2,176\\
Normalization / rotary base / embeddings & RMSNorm $+$ QK-norm / 100,000 / tied\\
Backbone parameters & 892,881,664\\
\midrule
Tag tables (emotion, genre, accomp., perf.) & $(6,4,6,5) \times 1{,}536$\\
Shared trunk (linear $+$ SiLU) & $1{,}536 \rightarrow 512$\\
Per-layer tag head, zero-initialized & $512 \rightarrow 6 \times 1{,}536$\\
Cross-attention heads $\times$ width, per layer & $16 \times 64$\\
Text-state width / null context tokens & 2,560 / 1\\
Added parameters / conditional total & 341,888,512 / 1,234,770,176\\
\bottomrule
\end{tabular}
\end{table}

\subsection{Autoregressive training}

Table~\ref{tab:ar-training} summarizes the training configuration of each stage. All stages use AdamW with weight decay 0.1, 1\% warm-up, cosine decay to 10\% of the peak learning rate, gradient clipping at 1.0, bfloat16 precision, seed 0, and four RTX A6000 GPUs. Token-budget batching allows 12,288 tokens per device microbatch. Conditional training drops the four perceptual tags and the imagery together with probability 0.10, replacing them with their null conditions, while key, tempo, and plan tokens remain in the event sequence.

\begin{table}[htbp]
\caption{Autoregressive training configuration by stage.}
\vspace{3mm}

\label{tab:ar-training}
\centering\small
\begin{tabular}{@{}lcccc@{}}
\toprule
 & {\bfseries Pretrain} & {\bfseries Large public} & {\bfseries Curated} & {\bfseries High-quality}\\
\midrule
Distinct pieces per pass & 816,682 & 369,553 & 50,130 & 9,849\\
Passes / training rows & 5 / 4,083,410 & 6 / 2,217,318 & 3 / 150,390 & 5 / 49,245\\
Validation rows & 1,518 & 1,500 & 1,000 & 100\\
Training tokens & 11.44B & 6.60B & 429.72M & 155.97M\\
Optimizer updates & 43,708 & 12,616 & 1,644 & 607\\
Peak learning rate & $2\!\times\!10^{-4}$ & $1.5\!\times\!10^{-4}$ & $5\!\times\!10^{-5}$ & $2.5\!\times\!10^{-5}$\\
Token budget per update / accumulation & 294,912 / 6 & 589,824 / 12 & 294,912 / 6 & 294,912 / 6\\
Condition dropout & -- & 0.10 & 0.10 & 0.10\\
\bottomrule
\end{tabular}
\end{table}

The first large-public pass orders pieces from sparse to complete perceptual labels, and later passes are shuffled. The high-quality training count follows the exclusions and split in Appendix~\ref{app:pipeline}. Architecture, conditioning, and augmentation remain fixed across the conditional stages.

\subsection{Latent representation and diffusion conditioning}
\label{app:dit-conditioning}

The MuseTok-based codec~\citep{musetok} maps each one-second window of the token sequence to a 128-dimensional posterior mean. Windows start every 0.5\,s, yielding two overlapping latents per second. Source rows are cropped to at most 190\,s, at a section boundary where possible, and the model has a 384-position canvas (192\,s). Batches are length-bucketed, and padding is excluded from attention and loss. 

Table~\ref{tab:dit-arch} specifies the diffusion transformer~\citep{peebles2023dit}. Its conditioning code sums a transformed timestep embedding with a transformed sum of label embeddings for the six tags and two flags. The completeness flag is set when a row ends at the end of its piece, and the interior-start flag when it begins after the opening. Per-position embeddings of section identity, within-section position, and absolute position carry the plan. Imagery uses the same frozen text states as AR, with a null token. Each block's AdaLN-Zero head supplies a shift, scale, and gate for each of its self-attention, cross-attention, and feed-forward branches. The modulation heads and the final output projection start at zero, so every branch is initially switched off.

\begin{table}[htbp]
\caption{Latent diffusion architecture with a 384-position canvas.}
\vspace{3mm}

\label{tab:dit-arch}
\centering\small
\begin{tabular}{@{}lc@{}}
\toprule
{\bfseries Parameter} & {\bfseries Value}\\
\midrule
Hidden width / layers / heads & 1,024 / 16 / 16\\
Feed-forward ratio (GELU) / latent width & $4\times$ / 128\\
Canvas positions / conditioning-code width & 384 / 512\\
AdaLN-Zero head output width per block & $9 \times 1{,}024$\\
Cross-attention context width & 2,560\\
Normalization & LayerNorm, no affine; QK-RMSNorm\\
Parameters & 397,558,400\\
\bottomrule
\end{tabular}
\end{table}

\subsection{Diffusion objective and training}

The rectified-flow objective minimizes the mean squared velocity error over non-padded positions. One timestep per sequence is drawn from a logit-normal distribution, $t=\sigma(u)$ with $u\sim\mathcal{N}(\mu,1)$, where $\mu=-0.5$ during pretraining and $-0.25$ during conditional training.

Classifier-free dropout~\citep{ho2022classifierfree} independently nulls three condition groups with probability 0.10 each: (1) perceptual tags and imagery, (2) key, tempo, and the completeness and interior-start flags, and (3) the section plan. These drops add to the missing or dropped conditions inherited from the corpus.

All stages use AdamW with $(\beta_1,\beta_2)=(0.9,0.95)$, weight decay 0.01, warm-up and cosine decay to 10\% of the peak learning rate, gradient clipping at 1.0, bfloat16 computation and gradient compression, and an exponential moving average with decay 0.9995. The moving-average weights initialize the next stage and are used for inference. Training uses two RTX 5090 GPUs with per-device batch size 24 and accumulation 8, for a global batch of 384. As in AR, a pass is one clean or augmented copy of the corpus and keep original pass last (Table~\ref{tab:dit-training}).

\begin{table}[htbp]
\caption{Diffusion training configuration. Curated (b) continues (a) with a restarted schedule. Pretraining rows per pass are listed as clean / augmented.}
\vspace{3mm}

\label{tab:dit-training}
\centering\small
\begin{tabular}{@{}lccccc@{}}
\toprule
 & \multirow{2}{*}{\bfseries Pretrain} & {\bfseries Large} & \multicolumn{2}{c}{\bfseries Curated} & {\bfseries High-}\\
\cmidrule(lr){4-5}
 &  & {\bfseries public} & {\bfseries (a)} & {\bfseries (b)} & {\bfseries quality}\\
\midrule
Steps & 120,000 & 120,000 & 24,000 & 36,000 & 9,000\\
Peak learning rate ($\times 10^{-5}$) & 10 & 10 & 5 & 3 & 2\\
Warm-up steps & 3,000 & 2,000 & 500 & 500 & 300\\
Timestep mean $\mu$ & $-0.5$ & $-0.25$ & $-0.25$ & $-0.25$ & $-0.25$\\
Passes visited & 2 & 3 & 3 & 3 & 3\\
Rows per pass & 816,633 / 816,641 & 369,552 & 50,129 & 50,129 & 9,849\\
Validation rows & 1,518 & 1,500 & 1,000 & 1,000 & 100\\
\bottomrule
\end{tabular}
\end{table}

\subsection{Inference settings}
\label{app:inference}

AR receives key and tempo tokens, the perceptual tags, and the imagery description. It first samples its section-plan prefix and then the performance events, with grammar masking to control the transition from plan to events. Sampling uses temperature 0.98, nucleus probability 0.98, no top-$k$ truncation, no guidance, and at most 4,094 new tokens.

DiT receives the section plan, latent length, and completeness flag of the reference performance. For 19 of the 100 prompts, the reference row is capped at the 4,096-token AR context, so DiT receives the plan and length of its opening, with the completeness flag unset. Fifty uniform Euler steps integrate from $t=1$ to $0$ with the guided velocity
\[
 v_{\varnothing}+3.0(v_c-v_{\varnothing}),
\]
where the null branch drops all three condition groups. Sampled latents are destandardized and decimated to integer-second offsets before grammar-constrained decoding in 20-second chunks, with a cap of 1,024 tokens per chunk and the end token masked because the duration is supplied.
\section{Evaluation and extended results}
\label{app:eval}

\subsection{Evaluation samples and rendering}

Objective results use all eight samples for each of the 100 held-out prompts, except that 12 failed MuseCoco decodes are excluded, leaving 788. The reference row contains the 100 paired performances. Appendix~\ref{app:inference} gives the sampling settings.

All systems and references set are rendered from MIDI with FluidSynth and the TimGM6mb soundfont at a shared gain. CLaMP~3 inputs are standardized to 960 ticks per beat and 120\,bpm while preserving performance timing. Distances use all 1,276 MAESTRO v3 and all 909 original POP909 performances. Audiobox scores are averaged over 10-second windows within each clip and then over clips.

\subsection{Baseline interfaces}
\label{app:baseline-interfaces}

Table~\ref{tab:baseline-interfaces} lists the inputs each baseline receives. MuseCoco accepts only the key mode, so its key agreement is scored on mode alone ($m$); quadrant-conditioned models are scored on quadrant-folded emotion ($q$). EMO-Disentangler receives no key but declares one in its first stage, so its key agreement is measured against that declared key ($\dagger$).

\begin{table}[htbp]
\caption{Baseline input mappings. Emotion quadrants map happy to Q1, angry/anxious to Q2, sad to Q3, and calm to Q4.}
\vspace{3mm}
\centering\small

\label{tab:baseline-interfaces}
\begin{tabular}{@{}lc@{}}
\toprule
{\bfseries System} & {\bfseries Received conditions and adaptation}\\
\midrule
\parbox[t]{0.29\linewidth}{\raggedright MuseCoco~\citep{musecoco2023}} & \parbox[t]{0.66\linewidth}{\raggedright Attribute-to-music stage: piano, emotion quadrant, genre, key mode, tempo class, 4/4 meter, and a length request over 60\,s. Native top-$k$ 15, temperature 1.0, and 512--2,560 tokens. Its tempo cutoffs (76/120\,bpm) differ from ours (75/115).}\\
\addlinespace
\parbox[t]{0.29\linewidth}{\raggedright text2midi~\citep{text2midi}, MIDI-LLM~\citep{wu2026midillm}} & \parbox[t]{0.66\linewidth}{\raggedright All six tags and imagery in a caption. Tempo classes become Adagio 65, Andante 95, or Allegro 135\,bpm. Decoding masks non-piano programs and drum notes.}\\
\addlinespace
\parbox[t]{0.29\linewidth}{\raggedright EMO-Disentangler~\citep{emo}} & \parbox[t]{0.66\linewidth}{\raggedright Emotion quadrant, decomposed into valence and arousal across its two stages. The first stage samples and declares its own key.}\\
\addlinespace
\parbox[t]{0.29\linewidth}{\raggedright Piano Transformer~\citep{lehmkuhl2025pianotransformer}} & \parbox[t]{0.66\linewidth}{\raggedright Genre prefix only.}\\
\addlinespace
\parbox[t]{0.29\linewidth}{\raggedright Aria~\citep{bradshawscaling}, Compose \& Embellish~\citep{wu2023compose}} & \parbox[t]{0.66\linewidth}{\raggedright Unconditional piano generation for quality evaluation.}\\
\bottomrule
\end{tabular}
\end{table}

Caption-conditioned baselines receive: \emph{``A \{emotion\} \{genre\} solo piano piece featuring only piano. Set in \{key\} with a 4/4 time signature, it moves at an \{tempo word\} tempo of \{bpm\} bpm, with \{accompaniment\} accompaniment, played \{performance\}. \{imagery description\}''}.

\subsection{Critics and interpretation}
\label{app:critics}

\paragraph{Key and tempo.}
Key and tempo are read back with the estimators of Appendix~\ref{app:estimators}. Reference key agreement is 1.000 by construction. Reference tempo labels come from audio beat tracking, and the MIDI estimator agrees with them on 0.820 of the reference pieces.

\paragraph{Tag critics.}
The tag critics apply class-balanced logistic regression to concatenations of CLaMP~3 (768-d) and Aria (512-d) embeddings. Their five-fold accuracies for emotion, genre, accompaniment, and performance style are 0.869, 0.852, 0.543, and 0.491. The lower accuracy for accompaniment and performance style matches their lower agreement on the reference set in Table~\ref{tab:control}.

\paragraph{External diagnostics.}
The external emotion critic (E4) is a ReLU MLP with 512 hidden units fitted on 1,078 EMOPIA clips~\citep{hung2021emopia}, with song-grouped five-fold accuracy 0.745. Zero-shot genre (G0) selects the most similar LAION-CLAP text embedding among ``A \{class\} solo piano piece.'' for the three genres. Table~\ref{tab:control-full} adds both diagnostics to the main comparison. AR leads E4 and G0 among the evaluated generators, whereas DiT leads on the genre tag critic, indicating that the ranking depends on the domain in which a critic was trained.

\paragraph{Full-prompt alignment.}
CLaMP~3 and CLAP report the mean music--text cosine similarity for ``A \{emotion\} \{genre\} solo piano piece with \{accompaniment\} accompaniment, played in a \{performance\} style. \{imagery description\}''. This caption evaluates tags and imagery jointly, whereas the listening study presents the imagery without tags.

\begin{table}[htbp]
\caption{Tag agreement with external diagnostics ($\uparrow$). E/G/A/P: emotion/genre/accompaniment/performance. E4: external emotion. G0: zero-shot genre. Ranking and the marks $m$, $q$, and $\dagger$ follow Table~\ref{tab:control}.}
\vspace{3mm}

\label{tab:control-full}
\centering
\resizebox{\linewidth}{!}{
\begin{tabular}{@{}lcccccccccc@{}}
\toprule
{\bfseries Model / stage} & {\bfseries Key} & {\bfseries Tempo} & {\bfseries E4} & {\bfseries E} & {\bfseries G0} & {\bfseries G} & {\bfseries A} & {\bfseries P} & {\bfseries CLaMP~3} & {\bfseries CLAP}\\
\midrule
Reference & 1.000 & 0.820 & 0.630 & 0.840 & 0.500 & 0.830 & 0.550 & 0.530 & 0.132 & 0.338\\
\midrule
MuseCoco & $0.582^m$ & 0.636 & 0.280 & $0.575^q$ & 0.315 & 0.386 & -- & -- & -- & --\\
text2midi & 0.181 & 0.456 & 0.268 & 0.276 & 0.330 & 0.380 & 0.205 & 0.270 & 0.054 & 0.273\\
MIDI-LLM & 0.340 & 0.713 & 0.281 & 0.285 & 0.391 & 0.370 & 0.201 & 0.226 & 0.059 & 0.292\\
EMO-Disentangler & $0.752^\dagger$ & -- & \underline{0.477} & $\mathbf{0.611^q}$ & -- & -- & -- & -- & -- & --\\
Piano Transformer & -- & -- & -- & -- & 0.375 & 0.458 & -- & -- & -- & --\\
\midrule
TASTE (AR) & \underline{0.767} & \underline{0.730} & \textbf{0.514} & \underline{0.585} & \textbf{0.444} & \underline{0.509} & \underline{0.296} & \underline{0.273} & \textbf{0.141} & \underline{0.335}\\
TASTE (DiT) & \textbf{0.993} & \textbf{0.809} & 0.406 & 0.584 & \underline{0.414} & \textbf{0.603} & \textbf{0.326} & \textbf{0.350} & \underline{0.123} & \textbf{0.347}\\
\bottomrule
\end{tabular}
}
\end{table}

\subsection{Temporal evaluation}
\label{app:structure-eval}

The retrained structure probe of Appendix~\ref{app:section-supervision} segments generations without any priors. Because we placed the boundaries based on the probe of the high-quality plans (Appendix~\ref{app:backward}), the reference boundary F1 is partly by construction. Evaluation uses boundary threshold 0.30, a 20\,s minimum peak spacing, grouping threshold 0.65, at most eight identities, merging of adjacent equal identities, and absorption of a final span shorter than five seconds. The complementary Foote-style novelty curve~\citep{foote2000novelty} uses sustain- and velocity-weighted chroma plus texture features at 2\,fps, cosine self-similarity, and a Gaussian-tapered checkerboard kernel with a 10-second half-width.

Boundary precision, recall, and F1 use greedy one-to-one matching within 5\,s, with counts pooled over the cohort. Planned boundaries within 2.5\,s of the last note-off are excluded. Contrast is the mean peak novelty within $\pm2.5$\,s of a piece's planned boundaries divided by its median interior novelty, and Tables~\ref{tab:structure} and~\ref{tab:structure-stages} report the cohort median. The random-position comparison uses equally many seeded interior positions with the same local-maximum rule. Pairwise F1 compares same-section frame pairs at 0.5-second resolution and is averaged over pieces. Single-section plans are excluded from boundary and contrast scores but retained for pairwise F1.

\subsection{Effect of staged training}
\label{app:stages}

Tables~\ref{tab:stages}, \ref{tab:stage-quality}, and~\ref{tab:structure-stages} report control, quality, and temporal structure after each conditional stage, together with an AR ablation that skips the large-public stage. The two refinement stages contribute different kinds of control. Curated supervision raises genre agreement in both families, most clearly for AR, but lowers emotion agreement and boundary F1. High-quality refinement then brings the largest gains in emotion agreement, CLaMP~3 alignment, and boundary F1, which all peak at this stage in both families, while AR's genre agreement partly falls back. Curated evidence thus mainly informs genre, whereas high-quality pairs improve affective, semantic, and temporal alignment.

Accompaniment and performance style develop less consistently across stages, in line with their less reliable labels. For DiT, the large-public stage already attains the highest tempo, accompaniment, and CLAP scores, which the later stages do not improve. In temporal structure, AR's recurrence agreement peaks at the curated stage, whereas DiT improves both boundary and recurrence agreement at the high-quality stage. Learning when to change and learning which material should recur thus respond differently to supervision, particularly when AR generates the plan along with the music.

Refinement also shifts the output distribution toward pop: symbolic distance to POP909 falls at each stage, and audio distance to MAESTRO rises from the first to the final stage. Production-quality and content-enjoyment scores are highest after the high-quality stage in both families.

The stage-wise rows cannot show how much the large-public stage itself contributes, since every later stage starts from it. To isolate it, we train AR from the pretrained model on the curated and then high-quality tiers with otherwise identical settings (\emph{w/o large public}). Compared with the full recipe, the final model changes only modestly. Emotion, genre, and accompaniment agreement drop by 0.019 to 0.044, CLaMP~3 alignment from 0.141 to 0.131, and boundary and pairwise F1 from 0.517 to 0.464 and from 0.608 to 0.592. Key, tempo, performance style, and the external critics stay level or slightly improve, and the quality metrics change little in either direction. Most of the control thus comes from the curated and high-quality tiers, while the large public tier adds a few points on emotion, genre, accompaniment, CLaMP~3 alignment, and temporal structure.

The ablation also ends with a higher validation loss on the high-quality tier (1.715 versus 1.695). In an informal check of a few samples, its outputs sounded structurally simpler and of lower quality, a gap that the automatic quality metrics may understate, as they did for DiT in Section~\ref{sec:listening}.

\begin{table}[htbp]
\caption{Control and alignment across training stages ($\uparrow$), with 800 samples per row. $+$ denotes continued training. \emph{w/o large public}: AR trained on the curated and high-quality tiers only, not ranked. Notation follows Table~\ref{tab:control-full}.}
\vspace{3mm}

\label{tab:stages}
\centering
\resizebox{\linewidth}{!}{
\begin{tabular}{@{}lcccccccccc@{}}
\toprule
{\bfseries Model / stage} & {\bfseries Key} & {\bfseries Tempo} & {\bfseries E4} & {\bfseries E} & {\bfseries G0} & {\bfseries G} & {\bfseries A} & {\bfseries P} & {\bfseries CLaMP~3} & {\bfseries CLAP}\\
\midrule
\multicolumn{11}{l}{\textit{AR}}\\
Large public & 0.642 & 0.731 & \underline{0.440} & 0.395 & 0.388 & 0.451 & 0.295 & 0.281 & 0.083 & 0.339\\
$+\,$Curated & 0.559 & 0.682 & 0.406 & 0.370 & \underline{0.434} & \textbf{0.615} & 0.236 & 0.300 & 0.090 & 0.332\\
$+\,$High-quality & 0.767 & 0.730 & \textbf{0.514} & \textbf{0.585} & \textbf{0.444} & 0.509 & 0.296 & 0.273 & \textbf{0.141} & 0.335\\
\emph{w/o large public} & 0.781 & 0.754 & 0.516 & 0.566 & 0.458 & 0.481 & 0.252 & 0.299 & 0.131 & 0.333\\
\midrule
\multicolumn{11}{l}{\textit{DiT}}\\
Large public & \underline{0.991} & \textbf{0.886} & 0.376 & 0.372 & 0.409 & 0.540 & \textbf{0.367} & \underline{0.349} & 0.087 & \textbf{0.351}\\
$+\,$Curated & 0.974 & 0.696 & 0.255 & 0.329 & 0.374 & 0.561 & 0.275 & 0.319 & 0.082 & 0.345\\
$+\,$High-quality & \textbf{0.993} & \underline{0.809} & 0.406 & \underline{0.584} & 0.414 & \underline{0.603} & \underline{0.326} & \textbf{0.350} & \underline{0.123} & \underline{0.347}\\
\bottomrule
\end{tabular}
}
\end{table}

\begin{table}[htbp]
\caption{Quality across training stages. Metrics follow Table~\ref{tab:quality}, and rows follow Table~\ref{tab:stages}.}
\vspace{3mm}

\label{tab:stage-quality}
\centering\small
\begin{tabular}{@{}lcccccc@{}}
\toprule
 & \multicolumn{2}{c}{\bfseries MAESTRO $\downarrow$} & \multicolumn{2}{c}{\bfseries POP909 $\downarrow$} & {\bfseries PQ $\uparrow$} & {\bfseries CE $\uparrow$}\\
\cmidrule(lr){2-3}\cmidrule(lr){4-5}
{\bfseries Model / stage} & {\bfseries V} & {\bfseries C} & {\bfseries V} & {\bfseries C} &  & \\
\midrule
\multicolumn{7}{l}{\textit{AR}}\\
Large public & \underline{0.210} & \underline{172.422} & 0.998 & 286.869 & 7.775 & 7.252\\
$+\,$Curated & 0.319 & 175.372 & 1.177 & 226.742 & 7.781 & 7.168\\
$+\,$High-quality & 0.426 & \textbf{170.697} & \textbf{0.605} & \underline{201.652} & \textbf{7.946} & \textbf{7.410}\\
\emph{w/o large public} & 0.445 & 165.549 & 0.592 & 206.130 & 7.947 & 7.416\\
\midrule
\multicolumn{7}{l}{\textit{DiT}}\\
Large public & 0.220 & 213.041 & 0.871 & 239.708 & 7.878 & 7.303\\
$+\,$Curated & \textbf{0.209} & 233.151 & 0.944 & 217.005 & 7.858 & 7.211\\
$+\,$High-quality & 0.309 & 239.704 & \underline{0.808} & \textbf{196.491} & \underline{7.945} & \underline{7.319}\\
\bottomrule
\end{tabular}
\end{table}

\begin{table}[htbp]
\caption{Temporal structure across training stages ($\uparrow$). P/R: boundary precision/recall. Parentheses give random-position contrast. Rows follow Table~\ref{tab:stages}.}
\vspace{3mm}

\label{tab:structure-stages}
\centering
\begin{tabular}{@{}lccccc@{}}
\toprule
 & \multicolumn{3}{c}{\bfseries Boundary @5\,s $\uparrow$} & {\bfseries Contrast $\uparrow$} & {\bfseries Pairwise}\\
\cmidrule(lr){2-4}
{\bfseries Model / stage} & {\bfseries P} & {\bfseries R} & {\bfseries F1} & {\bfseries (random)} & {\bfseries F1 $\uparrow$}\\
\midrule
Reference & 0.876 & 0.918 & 0.896 & 2.52 (1.72) & 0.769\\
\midrule
\multicolumn{6}{l}{\textit{AR}}\\
Large public & 0.384 & 0.301 & 0.338 & 1.88 (1.66) & 0.600\\
$+\,$Curated & 0.362 & 0.275 & 0.313 & 1.82 (1.65) & 0.623\\
$+\,$High-quality & 0.567 & 0.475 & 0.517 & 1.99 (1.69) & 0.608\\
\emph{w/o large public} & 0.518 & 0.420 & 0.464 & 1.93 (1.67) & 0.592\\
\midrule
\multicolumn{6}{l}{\textit{DiT}}\\
Large public & \underline{0.832} & \underline{0.770} & \underline{0.799} & \underline{3.02} (1.84) & \underline{0.721}\\
$+\,$Curated & 0.788 & 0.702 & 0.743 & 2.98 (1.84) & 0.707\\
$+\,$High-quality & \textbf{0.867} & \textbf{0.869} & \textbf{0.868} & \textbf{3.18} (1.78) & \textbf{0.775}\\
\bottomrule
\end{tabular}
\end{table}

\subsection{Listening protocol}
\label{app:listening}

The 24 listening prompts are sampled with a fixed seed under emotion quotas (5 angry, 5 anxious, 5 calm, 5 happy, 4 sad) from the prompts for which every system produced at least 10 seconds of audio. Clips use the opening 30 seconds of the first sample, with a one-second fade at the cut and peak normalization.

The 120 clips are distributed over eight questionnaire versions of 15 scored items, with three items per system and no prompt repeated within a version; an unscored practice item comes first. Each blind audio item first asks for overall quality from 1 (Bad) to 5 (Excellent) and then reveals two imagery descriptions for a forced choice.

\subsection{Listening results by musical training}
\label{app:training}

Table~\ref{tab:listening-training} splits the listening results by the participants' self-reported years of musical training. Listeners with more training rate every system lower, and training level correlates with a participant's mean quality rating (Spearman $\rho=-0.48$) but not with imagery accuracy ($\rho=-0.07$). Both groups still rate the reference highest and AR second, and both match TASTE outputs to their descriptions at the same rate (78.3\% and 77.8\% over AR and DiT). The imagery results thus hold for listeners with and without musical training.

\begin{table}[htbp]
\caption{Listening results by years of musical training: 10 participants with less than two years and 12 with two or more. Imagery accuracy is in percent; chance is 50\%.}
\vspace{3mm}
\label{tab:listening-training}
\centering\small
\begin{tabular}{@{}lcccc@{}}
\toprule
 & \multicolumn{2}{c}{\bfseries Quality (1--5)} & \multicolumn{2}{c}{\bfseries Imagery accuracy (\%)}\\
\cmidrule(lr){2-3}\cmidrule(lr){4-5}
{\bfseries System} & {\bfseries $<$2 years} & {\bfseries $\geq$2 years} & {\bfseries $<$2 years} & {\bfseries $\geq$2 years}\\
\midrule
Reference & 4.00 & 3.72 & 70.0 & 69.4\\
text2midi & 2.90 & 2.56 & 56.7 & 52.8\\
MIDI-LLM & 2.93 & 2.72 & 50.0 & 52.8\\
TASTE (AR) & 3.37 & 3.17 & 76.7 & 83.3\\
TASTE (DiT) & 2.97 & 2.53 & 80.0 & 72.2\\
\midrule
All systems & 3.23 & 2.94 & 66.7 & 66.1\\
\bottomrule
\end{tabular}
\end{table}

\end{document}